\documentclass[12pt]{article}
\usepackage{amsmath}
\usepackage{amssymb}
\usepackage{graphicx}
\begin{document}

\title{Analytical Methods for Measuring the Parameters of Interstellar
Gas Using the Data of Methanol Observations}

\author{S.~V.~Kalenskii\\Lebedev Physical Insitute of Russian Academy of Sciences, Astro Space Center\\Moscow, Russia\\
S.~Kurtz\\Institut de Radioastronom\'\i a y Astrof\'\i sica,\\ Universidad Nacional Autonoma de M\'exico\\Morelia, Michoac\'an, M\'exico}
\maketitle
%\email[E-mail:]{e-mail  €à¥á}
%\homepage[]{}% web-áâà ­šæ 
%\thanks{}% « £®€ à­®áâš
%\affiliation{¥áâ® à ¡®âë š/š«š  €à¥á  ¢â®à }
%\altaffiliation{â®à®¥ ¬¥áâ® à ¡®âë}
%\collaboration{ }
%\noaffiliation
%\lastcollaboration{ }
%\received{03.11.2015}% ®áâã¯š«  ¢ à¥€ ªæšî
%\revised{25.12.2015}% ®á«¥ €®à ¡®âªš

\begin{abstract}
We analyze methanol excitation in the absence of external radiation and 
consider LTE methods for probing interstellar gas. We show that 
rotation diagrams correctly estimate the gas kinetic temperature only if 
they are built from lines with the upper levels located in the same $K$-ladders,
such as the $J_0-J_{-1}E$ lines at 157~GHz, the $J_1-J_0E$ lines at 165~GHz
or the $J_2-J_1E$ lines at 25~GHz. The gas density should be no less
than $10^7$~cm$^{-3}$. Rotation diagrams built from lines with different $K$ 
values of the upper levels ($2_K-1_K$ at 96~GHz, $3_K-2_K$ at 145~GHz, 
or $5_K-4_K$ at 241~GHz) significantly underestimate the temperature 
but allow a density estimation. In addition, the diagrams based 
on the $2_K-1_K$ lines make possible methanol column density  estimates
within a factor of about 2--5. We suggest that rotation diagrams should be used in 
the following manner. First, one should build two rotation diagrams, one from 
the lines at 96, 145, or 241~GHz, and another from the lines at 157, 165, 
or 25~GHz. The former diagram is used to estimate the gas density. 
If the density is about $10^7$~cm$^{-3}$ or higher, the latter diagram reproduces 
the temperature fairly well. If the density is around $10^6$~cm$^{-3}$, 
the temperature obtained from the latter diagram should be multiplied by a 
factor of 1.5--2. If the density is about $10^5$~cm$^{-3}$ or lower, then
the latter diagram yields a temperature that is lower than the kinetic 
temperature by a factor of three or larger and should be used only as a lower 
limit on the kinetic temperature. Errors of methanol column density 
determined from the integrated intensity of a single line may be larger 
than an order of magnitude even when the gas temperature is well-known. 
However, if the $J_0-(J-1)_0E$ lines, as well as the $J_1-(J-1)_1A^{+}$ 
or $A^{-}$ lines are used, the relative error of the column density proves 
to be no larger than several units.
\end{abstract}

\section*{INTRODUCTION}
The methanol (CH$_3$OH) molecule is a slightly asymmetric top that has many
rotational transitions in the radio wave range. Its abundance in  
different types of interstellar clouds, from translucent clouds to dense
cores of Giant Molecular Clouds (GMC) and hot cores varies within
the range $\sim 10^{-9}-10^{-7}$~[1,2] and is high enough for methanol lines 
to be detectable in these objects. One of the features of methanol spectra
in the centimeter and millimeter wave ranges is the existence of several
series of lines so closely spaced in frequency that several lines from each 
series fall in the bandpass of any modern receiver and hence
can be observed together. Pointing and calibration errors in this case 
are the same for all the lines, and so the measured intensity ratios are not 
affected by these errors. 

Because of these properties methanol is an important tool for measuring 
the parameters of interstellar gas, especially since the cross-sections
for collisions of methanol with hydrogen were determined by quantum 
mechanical calculations~[3, 4]. This fact made it possible to build robust 
source models based on the results of Statistical Equilibrium (SE) 
calculations~[5--7].

Nevertheless, simple methods based on the Local Thermodynamic
Equilibrium (LTE) assumption are still widely used. Their use, however,
is not without problems. For example, rotation diagrams built from methanol 
lines often underestimate the kinetic 
temperature~(e.g.~[8,9]). The present paper elucidates the origin 
of these problems and shows how the observations should be carried out 
and analyzed in order to obtain correct results.

At its inception, this paper was intended to be a practical
manual describing the use of rotational diagrams and other analytical
methods as applied to methanol, describing the results that can be
obtained and the pitfalls that may be encountered. However,
the paper proved to be highly cumbersome and unreadable. Therefore the practical
part was shortened. Instead, the basic ideas on methanol excitation were
added, so that the readers could lean on them in the process of planning 
the observations and analyzing the results. Many important topics such as the spectroscopy
of methanol ($A$ and $E$ symmetry, selection rules etc.) are not
carefully developed, but only mentioned to the degree necessary for
understanding the remaining material. Interested readers can pursue
these topics in the book by Townes and Schawlow~[10], as well as
papers~[11,12].  The indispensable information from these references is
given in the next section. In addition, we do not consider the
influence of external radiation.  Thus, the conclusions and
recommendations from the present paper are valid only for those
sources where the influence of external radiation is not significant
--- such as dense cores in molecular clouds. Some short remarks on the
role of radiation are given in section~6.

\section{SPECTROSCOPY OF METHANOL}
\label{metspec:AstRus1608004Kalenskii}
The methanol molecule is a prolate, slightly asymmetric top (Fig.~1) and
has a large number of allowed transitions in the radio regime.
There are three types of methanol symmetry, designated $A$, $E1$, and $E2$.
However, the $J_KE1$ levels are degenerate with the $J_{-K}E2$ 
levels\footnote{$J$ is the total angular momentum quantum number; $K$ is
the quantum number of the angular momentum component along the symmetry 
axis of the CH$_3$ group.}. In addition, there are allowed 
dipole transitions between $E1$ and $E2$ states and vice versa. 
Therefore, one can correctly consider $E1$ and $E2$ states to be doubly 
degenerate states of $E$ symmetry where $K$ can take positive and negative 
values~[12]. Levels of methanol $A$ experience $K$-doubling (except for 
levels with $K=0$) and are labeled $A^{+}$ and $A^{-}$. Spin consideration shows 
that, as in the case of symmetric tops, there are no $J_0A^{-}$ methanol energy levels.

$A$ and $E$ symmetry is related to the alignment of nuclear spins in the hydrogen
atoms of the CH$_3$ group. In the case of methanol $A$ they are parallel and
in the case of methanol $E$ they are not. As nuclear spins interact weakly
with rotation and electric field, there are no allowed radiative
or collisional transitions between $A$ and $E$ methanol.
 
Selection rules for $A$-methanol have the form:

\begin{gather}\label{sela:AstRus1608004Kalenskii}
\Delta J = 0 \quad \Delta K = 0,{\pm} 1 \quad {\pm}
\leftrightarrow \mp \\
\nonumber \Delta J ={\pm} 1 \quad \Delta K = 0,{\pm} 1 \quad {\pm}
\leftrightarrow \pm
\end{gather}
Those for $E$-methanol are:
\begin{gather}
\label{sele:AstRus1608004Kalenskii}
\Delta J  =  0 \quad \Delta K ={\pm} 1 \\
\nonumber\Delta J  ={\pm} 1 \quad \Delta K = 0,{\pm} 1
\end{gather}

Level energies, transition frequencies, line intensities, and other
spectroscopic characteristics of methanol are presented in the JPL
(http://spec.jpl.nasa.gov), Cologne (http://www.astro.uni-koeln.de/cdms)
and Splatalogue (http://www.cv.nrao.edu/php/splat/) spectral line
catalogs. Level energies, transition frequencies, 
Einstein $A$ coefficients, and the collisional transition rates
can be found in the Leiden university database LAMDA
(http://home.strw.leidenuniv.nl/$\sim$moldata/). The energy level
diagrams for $A$ and $E$ methanol are presented in, e.g.,~Leurini et al.~[13].

\section{METHANOL EXCITATION}
\label{metexc:AstRus1608004Kalenskii}
Fig.~2 shows the energy levels of $A$ and $E$ methanol with $J\le10$. 
The ladders with the ground levels ($0_0A^{+}$ and $1_{-1}E$) are called 
the backbone ladders, and other ladders are called side ladders. An arrow 
downward from a level denotes the spontaneous transition with the largest 
Einstein $A$ coefficient for this level. The figure shows that such 
transitions are directed towards the backbone ladder. Therefore an excited
molecule after one or several spontaneous transitions appears in 
the backbone ladder, leading to an overpopulation of this ladder with 
respect to side ladders. This property of methanol excitation is responsible
for inversion in most of the Class I methanol
maser lines ($7_0-6_1A^{+}$, $4_{-1}-3_0E$, $5_{-1}-4_0E$, etc.), since the upper
levels of these lines belong to the backbone ladders, and the lower levels 
belong to side ladders~[12]. 

Consider the excitation of methanol in more detail.  We suggest that
the population of each level is mainly determined by the
fastest spontaneous transition out from this level. In
this section, the behavior of these transitions will be analyzed using
methods developed for two-level systems.  This approach is easy
and descriptive, but one can hardly consider that it is adequate for a
complex, multilevel system such as methanol. Therefore the
results of this section should be checked by SE
modeling, which will be done in the following section.

First, we consider the excitation of the $A^{+}$ levels, and in 
subsections~3.1 and 3.2 show how this differs in the cases of
$A^{-}$ levels and $E$  methanol levels.

Fig.~3 shows a segment of the $A$-methanol level configuration that 
includes several levels from the backbone ladder and the neighboring side 
ladders. Consider a system of three levels: $J_0A^{+}$ (denoted level 1), 
$J_1A^{+}$ (level 2), and $(J-1)_0A^{+}$ (level 3). As a specific example, 
we choose the $7_0A^{+}$ level as level 1; then level 2 will be $7_1A^{+}$ 
and level 3 will be $6_0A^{+}$. Note that according to selection 
rules (1), there are no allowed radiative transitions $J_1-J_0A^{+}$. 
The 2 to 3 transition has the largest Einstein A coefficient among all
spontaneous transitions from level 2 and hence, to the largest extent, determines
the lifetime of this level with respect to spontaneous emission.
For example, the lifetime of the $7_1A^{+}$ level, which is level 2 in our example
(and is the lower level of the well-known Class I maser transition $8_0-7_1A^{+}$
at 95~GHz), is above all determined
by the $7_1-6_0A^{+}$ transition; the A Einstein coefficient of this transition
is 1.6 $\times 10^{-3}$~s$^{-1}$, and the A Einstein coefficient of the second most
important downward transition from this level, $7_1-6_1A^{+}$, is $10\times$
smaller at \mbox{1.6 $\times 10^{-4}$~s$^{-1}$}.

All the Einstein coefficients and collisional constants used in
this paper are taken from the LAMDA database (see section~2).

For level 1, which belongs to the backbone ladder, there are no transitions
similar to the $2\rightarrow 3$ transition (see Fig.~2). The fastest spontaneous
transition from level 1 is the $1\rightarrow 3$ transition, whose 
Einstein $A$ coefficient depends on $J$ and hence, on the combination of levels
under consideration. In our example, the $1\rightarrow 3$ transition is 
the $7_0-6_0A^{+}$ transition with an Einstein A-coefficient of
 1.702 $\times 10^{-4}$~s$^{-1}$. On  average, the $1\rightarrow 3$ transitions are 
slower than the $2\rightarrow 3$ transitions by about an order
of magnitude. Therefore, the lifetimes of levels 1 with respect to
spontaneous emission are longer than the level 2 lifetimes by approximately 
an order of magnitude. In addition, the collisional rate 
coefficients for transitions between levels with the same $K$ quantum numbers
(in particular, for the $1\rightarrow 3$ transitions) are several times 
larger than those for transitions between levels with different $K$ numbers 
(in particular, for the $2\rightarrow 3$ transitions). Hence, there 
is a range of densities for which the level populations of the backbone ladder
are thermalized by collisions, while levels of the side ladders 
are underpopulated (by spontaneous transitions) with respect to the levels of 
the backbone ladder.  I.e., their populations are 
lower than they would be under LTE conditions with the given gas 
temperature\footnote{This
statement will be quantified using Eqs.~(7) and (9).}. 

If the levels in the backbone ladder up to level 1 are thermalized by 
collisions, i.e., the temperature  $T_{\textrm{rot}}$ that describes 
the ratios of their populations is approximaly equal to the kinetic 
temperature $T_{\textrm{kin}}$, then the population $n_1$ of this level can be 
expressed through the population of level 3 ($n_3$) with the Boltzmann equation:

\begin{gather}
\frac{n_1}{g_1}=\frac{n_3}{g_3} \exp \left({-}\frac{\Delta
E_{13}}{kT_{\textrm{kin}}}\right), \label{exc-eq1:AstRus1608004Kalenskii}
\end{gather}
where $\Delta E_{13}$ is the difference between the energies of levels
1 and 3, and $g_1$ and $g_3$ are  the statistical weights of these levels. The level~2 
population ($n_2$) can also be expressed via $n_3$:
\begin{gather}
\frac{n_2}{g_2}=\frac{n_3}{g_3} \exp \left({-}\frac{\Delta
E_{23}}{kT_{23}}\right), \label{exc-eq2:AstRus1608004Kalenskii}
\end{gather}
where $T_{23}$ is the excitation temperature of the $2\rightarrow 3$ transition. 
Combining Eqs.~(3) and~(4) one can find the population ratio for levels 1 and 2:
\begin{gather} \label{exc-eq5:AstRus1608004Kalenskii}
\frac{n_2}{g_2}=\frac{n_1}{g_1}\exp\left ({-}\frac{\Delta
E_{21}}{kT_{\textrm{kin}}}\right) \times \exp\left({-}\frac{\Delta E_{23}}{kT_{23}}+
\frac{\Delta E_{23}}{kT_{\textrm{kin}} }\right).
\end{gather}
Note that if the factor $\exp\Big(-\displaystyle{\frac{\Delta
E_{23}}{kT_{23}}}+{\displaystyle\frac{\Delta
E_{23}}{kT_{\textrm{kin}} }}\Big)$ is equal to unity, i.e., if
$T_{23}=T_{\textrm{kin}}$, then  according to Eq.~(5) the 
population ratio of levels 1 and 2 is determined by the kinetic temperature. 
In other words, if the $2\rightarrow 3$ transition is collisionally
thermalized, then the $2\rightarrow 1$ transition is also thermalized. As we show
below, $T_{23}$ is usually lower than $T_{\textrm{kin}}$ and hence
the factor is less than unity. Therefore, level 2 is  usually
underpopulated with respect to level 1; the lower $T_{23}$ is, the more
underpopulated   level 2 is.

Consider the relation between the populations of different methanol levels
and the gas parameters. If the population ratio of two levels
$u$ (the upper level) and $l$ (the lower level) is governed by collisional 
transitions and the spontaneous emission from the upper level, i.e.,
the $u\rightarrow l$ line is optically thin and the source is not illuminated
by strong external radiation,  then the population ratio can be found from 
 (e.g., Elitzur~[14])
\begin{gather} \label{exc-eq6:AstRus1608004Kalenskii}
\frac{n_u/g_u}{n_l/g_l} = 
\exp\left({-}\frac{\Delta
E_{ul}}{kT_{\textrm{kin}}}\right)\bigg/\left(1+\frac{A_{ul}}
{n_{\textrm{H}_2}C_{ul} } \right) = \\
\nonumber{}= \exp\left({-}\frac{\Delta
E_{ul}}{kT_{\textrm{kin}}}\right)\bigg/\left(1+\frac{n^{crit}}
{n_{\textrm{H}_2}}\right),
\end{gather}
where $C_{ul}$ is the collisional rate coefficient for the $u\rightarrow l$
transition and $n^{crit}$ is the critical density for this
transition; i.e., the density for which the condition 
$A_{ul}=n_{\textrm{H}_2}C_{ul}$ is fulfilled. From Eq.~(6) one can obtain:
\begin{gather}
\frac{n_1/g_1}{n_3/g_3}= \exp\left({-}\frac{\Delta
E_{13}}{kT_{\textrm{kin}}}\right)\bigg/
\left(1+\frac{n^{crit}_{13}}{n_{\textrm{H}_2}}\right).
\label{exc-eq7:AstRus1608004Kalenskii}
\end{gather}

The Einstein coefficients of spontaneous emission for the $1\rightarrow 3$
transitions increase with $J$ and for $J=7$ $A_{13}$ is 
$1.6\times 10^{-4}$~s$^{-1}$. The typical collisional rate coefficient
$C_{13}$  for a $1\rightarrow 3$ transition is about 
$4\times 10^{-11}$~cm$^3\cdot$s$^{-1}$. Hence, the critical density 
for a $1\rightarrow 3$ transition is about $5\times 10^6$~cm$^{-3}$ 
when $J=7$ and is lower when $J<7$. Therefore, when the density
is about $10^7$~cm$^{-3}$ the populations of the levels in the backbone 
ladder are collisionally thermalized up to $J=7$. At higher densities, higher 
levels will be thermalized as well.

For the $2\rightarrow 3$ transition one can find:
\begin{gather}
\frac{n_2/g_2}{n_3/g_3}= \exp\left({-}\frac{\Delta
E_{23}}{kT_{\textrm{kin}}}\right)\bigg/\left(1+\frac{n^{crit}_{23}}
{n_{\textrm{H}_2}}\right). \label{exc-eq8:AstRus1608004Kalenskii}
\end{gather}
Dividing  Eq.~(8) termwise by Eq.~(7) and taking into account that
$\Delta E_{23}=\Delta E_{13}+\Delta E_{21}$, one can obtain:
\begin{gather}
\frac{n_2/g_2}{n_1/g_1}=\exp\left({-}\frac{\Delta
E_{21}}{kT_{\textrm{kin}}}\right)
\left(\frac{n_{\textrm{H}_2}+n^{crit}_{13}}{n_{\textrm{H}_2}+n^{crit}_{23}}\right).
\label{exc-eq9:AstRus1608004Kalenskii}
\end{gather}

Because the critical densities of the $2\rightarrow 3$ transitions are higher than 
those of the $1\rightarrow 3$ transitions, Eq.~(9) shows that when 
$n_{\textrm{H}_2}\lesssim n^{crit}_{23}$ level 2 is underpopulated with 
respect to level 1; the lower the density the more underpopulated is  level 2.
The ratio of populations for levels 1 and 2 corresponds to the kinetic temperature
only when the density is several times higher than the critical density
of the $2\rightarrow 3$ transition. In the case of methanol $A$, typical values
of the Einstein coefficients $A_{23}$ are about $10^{-3}$~s$^{-1}$, while typical 
values of collisional rate coefficients $C_{23}$ are about 
$1-2\times 10^{-11}$~cm$^3\cdot$s$^{-1}$. Hence, the critical densities of
these transitions are about $10^8$~cm$^{-3}$. Such densities are higher than 
typical gas densities in molecular clouds, including dense cores. Therefore 
2 levels are usually underpopulated with respect to 1 levels.

Considering several 3-level systems $J_0A^{+}$, $J_1A^{+}$, and $(J-1)_0A^{+}$ 
with different values of $J$ one can see that the population ratios for 
the levels in the $K=1$ ladder (levels $J_1A^{+}$), within the framework of 
our model, are characterized by the same rotational temperature as the levels
located in the backbone ladder. Similarly, consideration of the system 
consisting of the levels $J_1A^{+}$, $(J-1)_1A^{+}$, and $J_2A^{+}$
shows that the $K=2$ ladder is underpopulated with respect to the $K=1$ ladder,
approximately to the same extent as the $K=1$ ladder is underpopulated with respect
to the backbone ladder.

\subsection{$A^{-}$ levels}
\label{aminus:AstRus1608004Kalenskii}
The analysis presented in the previous sections is not suitable for 
the $J_1A^{-}$ levels, because for these levels there are no allowed
downward transitions similar to the $2\rightarrow 3$ transitions. 
This happens because there are no $J_0A^{-}$ levels (see Sect.~1 and 
Fig.~2), and the  $J_1A^{-}{-}(J-1)_0A^{+}$ transitions are prohibited 
by selection rules~(1). The fastest downward radiative transitions from
the levels $J_1A^{-}$ (for $J\leq 9$) are the $J_1A^{-}{-}J_0A^{+}$, 
which depopulate the $J_1A^{-}$ levels, just as the $2\rightarrow 3$ 
transitions depopulate the $J_1A^{+}$ levels. However, the Einstein $A$ 
coefficients and the critical densities for the $J_1A^{-}{-}J_0A^{+}$ 
transitions are approximately an order of magnitude lower than those 
for the $2\rightarrow 3$ transitions. Therefore the $J_1A^{-}$, $J\leq 9$
levels, being underpopulated with respect to the backbone ladder levels, 
should be more highly populated than the $J_1A^{+}$ levels.

The $J_K A^{-}$, $K > 1$ levels are depopulated by the $J_{K}-(J-1)_{K-1}A^{-}$ 
transitions, just as the corresponding $A^{+}$ levels are depopulated
by the $J_{K}-(J-1)_{K-1}A^{+}$ transitions. Therefore the behavior of 
the $J_K A^{+}$~and~$J_K A^{-}$, $K > 1$ levels should be 
approximately the same.

\subsection{Methanol $E$}
\label{met-e:AstRus1608004Kalenskii}

The behavior of methanol $E$ is generally the same as that of methanol $A$:
both methanol $A$ and methanol $E$ have fast radiative transitions 
that depopulate the side ladders. However, there are some differences.
First, there are no $J_KE^{+}$ and $J_KE^{-}$ levels; instead, there are 
levels with positive and negative $K$ values (see Sect.~2 and Fig.~2).
The backbone ladder is the $K={-}1$ ladder (i.e., the sequence of 
the $J_{-1}E$ levels) and for the three-level analysis described above, 
one should choose the levels 1, 2, and 3 as the $J_{-1}E$, $J_0E$, 
and $(J-1)_{-1}E$ levels, respectively. The frequencies of the 
$2\rightarrow 3$ transitions prove to be lower than in the case of
methanol $A$ (see Fig.~2); as a result, the Einstein  coefficients $A_{23}$ 
of the former transitions are smaller than those of the latter 
transitions. For example, the $A_{23}$ value of a typical $2\rightarrow 3$ 
transition $5_0{-}4_{-1}E$, is only $1.4\times 10^{-4}$~s$^{-1}$, 
 which is an order of magnitude lower than that of the typical $A_{23}$ 
value of ${\sim} 10^{-3}$~s$^{-1}$ in the case of methanol $A$.
Methanol $E$ collision rate coefficients $C_{23}$ are also lower than 
those of methanol $A$, but to a lesser extent: the ratios between 
the $C_{23}$ values of $A$ and $E$ methanol are about 2--3. 
As a consequence, Eq.~(9) shows that the side ladders of methanol $E$
$K=0$, $K=1$ etc. are depopulated to a lesser extent than the side
ladders of methanol $A$ at the same temperature and density.

Unlike the case of  $A$ methanol, the $2\rightarrow 1$ transitions of  $E$
methanol {\it are} allowed by selection rules. However, the $2\rightarrow 3$ and 
$1\rightarrow 3$ transitions are much faster than the $2\rightarrow 1$ transitions. 
As a result, the ratio between the populations of levels 1 and 2, as in the case 
$A$ of methanol, is determined mostly by the $2\rightarrow 3$ and $1\rightarrow 3$ 
transitions, rather than by the $2\rightarrow 1$ transitions.

Strictly speaking, the three-level analysis described in section 2
is not applicable to the levels $J_{-2}E$ and others located left
of the backbone ladder in the energy level diagram, since the depopulation 
of these levels is not dominated by {\emph{single}} radiative 
transitions. For example, the fastest downward transition from the level 
$5_{-2}E$ is the  $5_{-2}{-}4_{-1}E$ transition, denoted by an arrow
in Fig.~2. Its Einstein $A$ coeffecient is $1.3\times 10^{-3}$~s$^{-1}$. 
The second important downward transition from this level is 
the $5_{-2}-5_{-1}E$ transition, whose $A$ coefficient is
$0.4\times 10^{-3}$~s$^{-1}$; i.e., it is less than the $A$ coeffecient
of the former transition by a factor of $\sim$3. The same is true
for other levels located to the left of the backbone
ladder, showing that the three-level approach is not applicable
for them.

The Einstein coefficients for the transitions that
depopulate $E$-methanol levels to the left of the backbone ladder
are about an order of magnitude larger than the Einstein
coefficients for the $2\rightarrow 3$ transitions to the right
of the backbone ladder. This means that the depopulation of the ladders 
to the left of the backbone happens at higher rates
and their populations are significantly lower than those 
of the ladders to the right of the backbone ladder. This
conclusion is confirmed by SE calculations (see section~4). As 
the populations of the $J_KE$,~$K\leq -2$ levels are low, 
the $J_K{-}(J-1)_KE$,~$K\leq -2$ lines are significantly weaker
than the $J_K{-}(J-1)_KE$,~$K\geq 0$ lines, which can affect 
the rotation diagram analysis.

\subsection{The effect of the microwave background}
\label{mwb:AstRus1608004Kalenskii}

As noted in Section 1, in this paper we neglect the effects of
external radiation radiation fields.
However, the microwave background exists everywhere
and one should understand under what conditions 
it can significantly affect the results of methanol observations. 
Therefore in this section we briefly discuss the role of 
the microwave background.

Fig.~4 shows an example of the density dependence of the excitation 
temperature $T^{21}_{\textrm{ex}}$ of a typical $2\rightarrow 1$ 
transition in the case of $E$ methanol, calculated using Eq.~9 
for different kinetic temperatures. For this example we chose 
the system of levels $5_{-1}E$ (level 1), $5_0E$ (level 2), 
and $4_{-1}E$ (level 3), i.e., the $2\rightarrow 1$ transition
is a well-known transition $5_{-1}-5_0E$. Note that any other 
$J_{-1}{-}J_0E$ transition may be the $2\rightarrow 1$ transition in 
a relevant system of the~1, 2, and 3 levels. The figure shows that for 
densities below $10^7$~cm$^{-3}$ the dependence of $T^{21}_{\textrm{ex}}$ 
on the kinetic temperature is fairly weak. $T^{21}_{\textrm{ex}}$ 
decreases with decreasing density and at densities below 
${\sim} 10^6$~cm$^{-3}$ it becomes close to the microwave 
background temperature, making the  $J_{-1}-J_0E$ lines weak or 
invisible even if the column density of methanol is high. 

According to Eqs.~(7) and~(8), which neglect radiation, the excitation 
temperatures of the $2\rightarrow 3$ and $1\rightarrow 3$ transitions 
tend to zero with density tending to zero. In fact, the role
of collisional transitions diminishes as the density
decreases, and at densities about $10^5$~cm$^{-3}$ 
the population ratios $n_1/n_3$ and $n_2/n_3$ are 
governed by the microwave background jointly with
collisions. If the density is about $10^4$~cm$^{-3}$
or lower, they are almost completely governed by
the microwave background. As a result, the excitation temperatures 
of the $2\rightarrow 3$ and $1\rightarrow 3$ transitions 
become close to the microwave background temperature 
and these lines become invisible too. 

Thus, the role of the microwave background increases with decreasing
density and becomes significant starting from several units
$\times 10^5$~cm$^{-3}$, which is fairly common in dense cores
of molecular clouds.

\section{RESULTS OF SE CALCULATIONS}
\label{secsec:AstRus1608004Kalenskii}

Methanol has a complex system of energy levels. Therefore, the conclusions 
of the previous section, based on assumptions valid for two-level systems,
should be tested by SE calculations. 

Figure~5 shows the results of LVG modeling, obtained with the RADEX 
software~[15]. The left column demonstrates the results for 
methanol~$A$, the right, for methanol~$E$. The $X$-axes plot 
the level energies, divided by the Boltzmann constant, $E/k$,
while the $Y$-axes plot the decimal logarithms of the level populations 
divided by their statistical weights, $\lg n/g$.  Levels from 
the same $K$-ladders are connected by solid or dashed lines.
Each of these lines can be considered as a rotation diagram 
(see~section 5.1), built with methanol levels of 
the same $K$ value. The figure shows that the main conclusion 
of the previous section --- that the side ladders are 
underpopulated with respect to the backbone ladders up to a density 
of about $10^8$~cm$^{-3}$ --- is confirmed by the SE calculations. 
In addition, one can see that the $\lg n/g$ values for levels
within the same ladder can be satisfactorily fitted with a straight 
line, and hence correspond to a common rotational temperature\footnote{This 
rotational temperature becomes approximately equal to the kinetic temperature 
at a density of order a few $\times 10^7$~cm$^{-3}$ 
(see below)}.

Figure 5 also shows that the behavior of ratios 
between the populations of different ladders is more complex than 
predicted by the simple model of the previous 
section. The difference occurs because the simple model
ignores some important factors that affect the level populations,
principally the transitions into levels 1 and 2 from higher-lying levels. 
For example, the SE calculations show that the population 
of the $K=2$ ladder of $E$ methanol is higher than that of the $K=1$ 
ladder, which is closer to the backbone ladder. The $K=2$ ladder is 
overpopulated with respect to the $K=1$ ladder because of the $J_3{-}(J-1)_2E$ 
transitions, which are faster than the $J_2{-}(J-1)_1E$ transitions 
(see, e.g.,~[12]), which empty the $K=2$ ladder. The overpopulation 
of the $K=2$ ladder results in masing of the $J_2{-}J_1E$ lines 
at 25 GHz.

In addition, at low densities ($n_{\textrm{H}_2}\approx
10^4$~cm$^{-3}$) the difference
between the populations of the $K=1$ and 2 ladders of methanol $A$ proves 
to be very small, significantly less than the difference between 
the populations of the $K=0$ (backbone) and $K=1$ ladders.
Note that at such low densities the level populations are not
determined by collisions~(see~section~3.3), and the model
developed in section~3 is not valid.

\section{\bf IMPLICATIONS FOR METHANOL OBSERVATIONS}
\label{analysis:AstRus1608004Kalenskii}

\subsection{Rotation diagrams}
\label{rott:AstRus1608004Kalenskii}

One of the most widespread methods for determining gas temperature 
is the rotation diagram method (e.g.,~[16]). Molecular column density 
in the upper level of an optically thin line, divided by the statistical 
weight of this level ($N_u/g_u$), can be determined with the equation
\begin{gather}
\frac{N_u}{g_u} = \frac{3kW}{8\pi^3\nu_0S\mu^2}, \label{eqsl:AstRus1608004Kalenskii}
\end{gather}
where $W=\int T_R dV$ is the integrated intensity of the line, $\nu_0$
is the central frequency of the line, and $S\mu^2$ is the product of the line 
strength and the squared permanent dipole moment. Equation~(10) 
is valid when the excitation temperature of the transition is much 
higher than the microwave background brightness temperature, which is 
not always true in molecular clouds (see subsection~3.3).
In LTE at a temperature $T_{\textrm{rot}}$ the level populations 
are distributed according to
\begin{gather}
\ln\frac{N_u}{g_u} = \ln\frac{3kW}{8\pi^3\nu_0S\mu^2}
= \ln\frac{N}{Q_{\textrm{rot}}}
{-}\frac{E_u}{kT_{\textrm{rot}}}, \label{eqrd:AstRus1608004Kalenskii}
\end{gather}
which follows from the Boltzmann equation. Here $N$ is the molecule column
density and $Q$ is the rotational partition function.

Suppose that we observe several lines of the same molecule. We can
 calculate  ${N_u}/{g_u}$  for each line and plot 
the dependence of $\ln({N_u}/{g_u})$ on $E_u/k$; the resulting plot is called 
a rotation diagram. Equation~(11) shows that under the LTE assumption
the points (which correspond to different transitions), lie on a single straight 
line whose slope is inversely proportional to $-T_{\textrm{rot}}$ and whose
intercept equals $\ln(N/Q_{rot})$. Thus, rotation diagrams make it 
possible to find both $T_{\textrm{rot}}$ and $N$.

Observations generally do {\it not} yield the brightness temperature $T_R$,
but rather its product with a (possibly unknown) beam filling factor, $T_R\cdot ff$.
One can easily show that if $ff$ is the same for all 
the observed lines then it will have no effect on the derived rotational temperature.
The molecular column density found by the rotation diagram, on the other hand,
will be the product $N\cdot ff$; i.e.,  the RD method yields the {\it beam-averaged} molecular column 
density.

From the above considerations it is clear that the rotational temperature derived from 
a {\emph{correctly built}} rotation diagram is the temperature that 
best fits the population ratios of a given set of energy 
levels. Here, we also consider {\emph{incorrectly
built}} rotation diagrams, which formally yield a
``rotational temperature'', but one which does {\it not} match the population ratios,
and in this sense is incorrect. Moreover, we discuss not only rotation 
diagrams, but other methods for analyzing optically thick lines. 
In all these cases we use ``rotational temperature'', or ``correct rotational
temperature'' to refer to the parameter that best describes the population ratios.

Rotation diagrams are often built using the methanol lines $2_K{-}1_K$ at 
96~GHz, $3_K{-}2_K$ at 145~GHz, or $5_K{-}4_K$ at 241~GHz; i.e., lines
with the same $J$ but different $K$ values of the upper levels. We call 
such diagrams type I rotation diagrams (RDIs), and the rotational 
temperatures derived with these diagrams we call type I rotational temperatures 
(RTIs).  Alternatively, rotation diagrams can be built from  lines 
whose upper levels belong to the same $K$-ladders. Suitable lines for 
this purpose are the well-known lines $J_0{-}J_{-1}E$ (157~GHz), 
$J_1{-}J_0E$ (166~GHz), and $J_2{-}J_1E$ (25~GHz). 
We call such diagrams type II rotation diagrams (RDIIs), and 
the rotational temperatures derived with these diagrams we call type II 
rotational temperatures (RTIIs). 

Below we consider the properties of both RDIs and RDIIs and show that 
they are fairly different. Rotation diagrams of a general kind 
(i.e., when {\it both} the $J$ and $K$ values of the applied levels are different)
will be analyzed in a subsequent paper.

\subsection{Rotation diagrams: quantitative considerations}
\label{rotqc:AstRus1608004Kalenskii}

Consider the construction of type I rotation diagrams. Examples of 
the $2_K{-}1_K$, $3_K{-}2_K$, and $5_K{-}4_K$ lines at 96, 145, and 
241~GHz, respectively, observed in molecular clouds,
are presented in Fig.~6. Rotation diagrams built from the 96-GHz 
lines are denoted RD96, those built from the 145-GHz lines, RD145, and
those built from the 241-GHz lines, RD241. Four lines are usually observed
at 96~GHz, namely, the  $2_{-1}{-}1_{-1}E$, $2_0{-}1_0A^{+}$,
$2_0{-}1_0E$, and ${2_1{-}1_1E}$ lines.  In cold, dark clouds the 
latter two lines may be very weak and undetectable (in a realistic observing 
time) due to the low temperatures and densities. 

If a wideband ($>2$~GHz) receiver is used, the $2_1{-}1_1A^{+}$ and $2_1-1_1A^{-}$
lines will also fall in the passband. The former line is 
about a gigahertz lower in frequency while the latter line is 
about a gigahertz higher than the four ``main'' lines. At 145~GHz 
usually the  $3_0{-}2_0E$, $3_{-1}{-}2_{-1}E$, $3_0{-}2_0A^{+}$,
and $3_1{-}2_1E$ lines are detected, as well as a blend of very closely 
spaced $3_2{-}2_2E$ and $3_{-2}{-}2_{-2}E$ lines. At 241~GHz 
the $5_0{-}4_0E$, $5_{-1}{-}4_{-1}E$, $5_0{-}4_0A^{+}$, $5_1{-}4_1E$ lines,
and the blended $5_2{-}4_2E$ and $5_{-2}{-}4_{-2}E$ lines, are 
usually observed. The blends of the $3_2{-}2_2E$ and $3_{-2}{-}2_{-2}E$ lines
and also the $5_2{-}4_2E$ and $5_{-2}{-}4_{-2}E$ lines are 
often used when building rotation diagrams, ascribing the integrated 
intensities of the spectral features
to the $3_2{-}2_2E$ and $5_2{-}4_2E$ lines, respectively.  However, we show below
that this practice can lead to incorrect rotation
temperatures (i.e., rotational temperatures that do not correspond to the 
population ratios of the $3_K$ or $5_K$ levels). When the main lines have
a sufficiently high
signal-to-noise ratio, one can observe other weak 
spectral features at 145 and 241~GHz that belong to the $3_K{-}2_K$ and $5_K{-}4_K$ line 
series. However, essentially all such features are blends of two or more 
lines. These features are rarely used for building rotation diagrams and 
will not be considered here.

Type I rotational temperatures are usually about 5--15~K, which is
significantly lower than the  gas temperatures obtained with other 
methods~[8,9]. It is sometimes suggested that the low rotation 
temperatures are a result of subthermal excitation of 
the $J_K{-}(J-1)_K$ lines; based on this suggestion,
the critical densities for these lines are treated as upper limits on 
the gas densities. In fact, the low rotational temperatures 
appear because RDIs are built from transitions with the same $J$
values of the upper levels, but belonging to different $K$-ladders.
Fig.~2 shows that in this case the greater the energy of the level,
the further the level will be from the backbone ladder. As shown in
Sections~3 and 4, this means that the higher energy levels will be more
severely underpopulated.  Using Eq.~(11) one can 
easily show that in this case the rotational temperature will be lower 
than the kinetic temperature. 
%It should be emphasized
%that the low rotational temperatures represent the real ratios of populations
%for the used sample of energy levels. 

As shown in Sections~3 and 4, the side ladders are underpopulated when
the gas density is lower than, or on the order of, the critical
density of the $2\rightarrow 3$ transitions. Therefore, in the case of 
a low RTI value,  one should use the critical density of these
transitions (as an upper limit to the gas density) rather than the
critical density of the $J_K{-}(J-1)_K$ transitions.  The former
critical densities are a few $\times 10^7$~cm$^{-3}{-}10^8$~cm$^{-3}$
(see Section~3).

Based on the results of sections~3 and 4, one can conclude that, in order 
to determine the kinetic temperature, rotation diagrams should be 
built from the lines whose upper levels belong to the same $K$-ladders;
i.e., one should use type II rotation diagrams. Suitable lines for this 
purpose are the well-known lines $J_0{-}J_{-1}E$ ($157$~GHz), 
$J_1{-}J_0E$ ($166$~GHz), and $J_2{-}J_1E$ ($25$~GHz). The levels giving
rise to the 25~GHz lines can suffer a population inversion (see section~4); 
however, at low optical depths inverted lines are as suitable for creating 
rotation diagrams as are ordinary thermal lines. 

\subsection{Rotation diagrams: SE results}
\label{rotseq:AstRus1608004Kalenskii}

To explore the quantitative properties of RDIs and RDIIs, we applied the LVG 
method and computed model brightness temperatures of methanol lines at 96,
145, and 241~GHz, and also at 157, 166, and 25~GHz.
The models were computed with the RADEX software~[15] for
kinetic temperatures 20--100~K and different gas densities and methanol
specific column densities. We assume the $A$ and $E$ 
methanol abundances to be equal.  

For each model we constructed both type I and type II rotation diagrams.
The level energies were counted off from the ground states
($0_0A^{+}$ in the case of methanol $A$ and $1_{-1}E$ in the case of 
methanol $E$). The results are shown in Table~1 and Fig.~7. Columns 2,3, 
and 4--7 of Table~1, right of the slashes, present type I 
rotational temperatures. Columns 3, 5, 7 show the ratios between 
the column densities derived from the RTIs and the model column densities.
Columns 8--13 present the same sets of parameters derived from RTIIs. 
The table shows the results for 50~K only. However, all the results, 
described below are valid for all kinetic temperatures within 
the range 20--100~K.

Based on Table~1 and Fig.~7 one can make the following conclusions 
concerning RTI. At low densities ($10^4-10^6$~cm$^{-3}$) RTI increases
with density but remains much lower than the kinetic temperature, falling 
in the range 3--8~K. At higher densities, RTI increases but remains 
significantly lower 
than the kinetic temperature even at densities  $10^7{-}10^8$~cm$^{-3}$.
Fig.~7 shows that for methanol column densities lower than  
${\sim} 10^{15}$~cm$^{-2}$/(km s$^{-1}$),\footnote{When 
the specific column density becomes as high as 
${\sim} 10^{15}$~cm$^{-2}$/(km s$^{-1}$) the lines used for building RDIs
become optically thick; i.e., one of the conditions for the applicability 
of rotation diagrams is violated.}  RT145 is largely independent of
both kinetic temperature and methanol column density; the same is true 
for RT96 and RT241. This behavior of type I rotational temperature is 
a consequence of the same features of methanol excitation as the property
previously found by Leurini et al.~[6]: the intensity ratios
of several $J_K{-}(J-1)_K$ and $J_{K-1}{-}(J-1)_{K-1}$ lines within the range of 
15--100~K depend mainly on density and can be used to estimate it.

Thus, type I rotation diagrams underestimate the kinetic temperature, as 
expected based on the results of sections~3 and 4. However, Fig.~7 and 
Table~1 show that they can be used to estimate {\emph{the gas density}}. 
At a density about $10^4{-}10^5$~cm$^{-3}$ RTI vary from 2--5~K, 
regardless of the kinetic temperature. At a density about 
$10^6$~cm$^{-3}$ RTI falls in the range 6--8~K; RTI of the order of 11~K 
or higher indicates densities of at least $10^7$~cm$^{-3}$.

The blend of the $3_2{-}2_2E$ and $3_{-2}{-}2_{-2}E$ lines is often used 
for building RD145s while the $5_2{-}4_2E$ and $5_{-2}{-}4_{-2}E$ blend
is used for RD241s (see subsection~5.2).  To investigate 
how the use of these blends affects RT145 and RT241 we
constructed model rotation diagrams including these blends.
The blended emission of the $5_2{-}4_2E$ and $5_{-2}{-}4_{-2}E$ lines
was attributed to the $5_2{-}4_2E$ line. Similarly, the blended emission
of the $3_2{-}2_2E$ and $3_{-2}{-}2_{-2}E$ lines was 
attributed to the $3_2{-}2_2E$ line\footnote{SE calculations showed that 
the $3_2{-}2_2E$ and $5_2{-}4_2E$ lines dominate the corresponding blends 
over most of the range of densities typical for molecular clouds.}. 
The results are presented in Columns 4--7 of Table~1 to the left of the slashes. 
The rotational temperatures at both 145 and 241 GHz at densities less
than $10^6$~cm$^{-3}$ are slightly higher than the correct 
values shown to the right of the slashes. However, at higher densities the derived 
rotational temperature increases rapidly with density, and at
about $10^8$~cm$^{-3}$ it can exceed the gas kinetic temperature. 
This happens because at high density the contribution from 
the $3_{-2}{-}2_{-2}E$ and $5_{-2}{-}4_{-2}E$ lines to the brightness 
temperatures of the corresponding blends becomes as large as 30\%--40\%. 
As a result, the column densities of methanol in the $3_2E$ and $5_2E$ 
levels are overestimated, which in turn leads to a significant 
overestimate of the temperature. Thus, the rotational temperatures 
derived using these blends generally do not correctly describe the ratios 
of the $3_KE$ or $5_KE$ level populations; they do not reproduce 
the kinetic temperature and cannot be used for the density estimation. 
Therefore these blends should not be used for building rotation diagrams.

Columns 8--13 of Table~1 exhibit rotational temperatures and methanol
column densities derived from type II rotation diagrams, built using 
the $J_0{-}J_{-1}E$ lines (RD157), $J_1{-}J_0E$ lines (RD165), and 
$J_2{-}J_1E$ lines (RD25). The table shows that rotation diagrams
built using different line series yield virtually the same rotational 
temperature. At densities below $10^6$~cm$^{-3}$ the rotational 
temperature is significantly lower than the gas kinetic temperature, but
higher than type I rotational temperatures for the same density.
When densities reach $10^7{-}10^8$~cm$^{-3}$ the type II rotational 
temperature coincides with the gas kinetic temperature. 

Our results show that to estimate the kinetic temperature one 
should build two rotation diagrams: RDI {\it and} RDII.  The RDI serves to
estimate the gas density\footnote{To evaluate the gas density, 
instead of building RDIs one can use the dependences of the ratios 
of different $J_K{-}(J-1)_K$ and $J_{K-1}{-}(J-1)_{K-1}$ line brightness
temperatures on density, presented in Figs.~4 and 5 from 
Leurini et al.~[6]}. Knowledge of the density makes it possible 
to estimate the temperature from the RDII. If the density is no higher than 
$10^5$~cm$^{-3}$, then the RTII is a lower limit to 
the gas kinetic temperature. When the density is about $10^6$~cm$^{-3}$, 
the RTII is lower than the kinetic temperature by a factor of 1.5--2. 
If the density is about $10^7$~cm$^{-3}$ or higher, then the RTII represents
the kinetic temperature with a high accuracy.

Methanol column density, as determined from a rotation diagram, can also be
erroneous and should be used with caution. When it is derived using 
RD96 or RD145, it may differ from the true value by a factor of 2--5, 
depending on density. If the column density is determined using RD241 
or any RDII, the accuracy will strongly depend on the source density. When 
the density is lower than $10^6$~cm$^{-3}$, the accuracy is very low: 
the ratio between the true and derived values may vary from 0.01--0.003
for RDII to ${\sim} 150$ for RD241. When the density is  about 
$10^6$~cm$^{-3}$, RDIIs can determine methanol column 
density with an accuracy of a factor of about five, and RD241s
overestimate it by a factor of about 15--20. When the density is 
near $10^7$~cm$^{-3}$ or higher, RD241s determine column density within
a factor of about 1.5,  RD25 overestimates it by a factor of 1.5--2, 
RD157s and RD165s overestimate it by a factor of 1.5--2.

In most of calculated models the 157~GHz lines were stronger than 
the 165 or 25~GHz lines and from this standpoint they are more suitable 
for building RDIIs.  Nevertheless, they do have some disadvantages. 
First, at low volume density and methanol column density 
the excitation temperatures of these 
lines are close to the microwave background temperature and so the lines 
are invisible (see section~3.3 and Table~1). Furthermore,
when the linewidth is about 2.5~km/s or broader, the $1_0{-}1_{-1}E$ and 
$3_0{-}3_{-1}$ lines are heavily blended, and when the linewidth 
is 5~km~s$^{-1}$ or higher, {\it three} lines, $1_0{-}1_{-1}E$, 
$3_0{-}3_{-1}E$, and $2_0{-}2_{-1}E$, are blended.
To obtain individual line parameters, one should assume that all 
the lines have the same radial velocities and/or linewidths. But
if the line profiles are not Gaussian, this approach is invalid,
and one cannot determine the line parameters individually. This 
disadvantage is not too serious, because in many clouds it is possible 
to detect four or more {\emph{non-blended}} lines of this series 
($J\ge 4$). A more important drawback is that when the specific 
column density of methanol becomes too high (several
$\times~10^{14}$~cm$^{-2}/$(km~s$^{-1}$ or higher) the 157~GHz lines
become optically thick. Such column densities are fairly common in 
the dense cores of molecular clouds. Fortunately, one can easily see if
a rotational diagram has been built from optically thick $J_0{-}J_{-1}E$ 
lines just by looking at its shape. When the lines are optically thick, 
the points
in the plot are located along an arc rather than along a straight line 
(see Fig.~8, lower diagram). In this case, the derived rotational 
temperature will depend on the particular sample of the lines used 
and can be either 
higher or lower than the true rotational temperature. For example, if 
we determine the rotational temperature using the optically thick 
$J_0{-}J_{-1}E$ lines $J=4{-}7$ (the four right-most points in the lower 
diagram in Fig.~8), the value obtained will be higher than the true temperature. 
If we use the $J=1{-}4$ lines (the four left-most points of
the same diagram) the derived value will be lower than the correct temperature. 
Thus, if the $J_0{-}J_{-1}E$ lines are optically thick, one should
build rotation diagrams from the $J_1{-}J_0E$ lines at 165~GHz;
but if the column density of methanol is higher than 
$\sim 10^{15}$~cm$^{-2}$, these lines will be optically thick as well. 

 \subsection{Modifications of the rotation diagram method, applicable 
in the case of optically thick lines}
\label{thick:AstRus1608004Kalenskii}

\textbf{5.4.1. Iterative method.} To determine gas parameters from 
optically thick lines, several modifications to the rotation
diagram method have been developed. One of these is an iterative 
procedure, applied, e.g., by Remijan et al.~[17] to the analysis 
of methyl cyanide (CH$_3$CN) observations. The procedure
is based on a method to account for the line optical
depth $\tau$.  In particular, the molecular column density in a level $u$, derived 
from Eq.~(10) (which presumes the line $u-l$ to be optically thin), can be 
multiplied by a correction factor $C_\tau=\tau /(1-e^{-\tau})$~[18].
In the first step of the iterative procedure it is assumed 
that the lines are optically thin, so initially the rotational temperature and
column density are determined from the usual rotation diagram. Based
on these values, and assuming LTE, one can derive each line's optical 
depth and correction factor $C_\tau$, and thus calculate the corrected column 
densities of the upper levels of the lines.  Then, using 
the corrected level column densities, one finds the next approximation for 
the rotational temperature and molecular column density. 
The procedure is repeated until convergence is achieved.

Unfortunately, to calculate the optical depths and the correction factors 
$C_\tau$ one must know the beam filling factor $ff$, which is often not the case.

\textbf{5.4.2. Population diagram method.} Rotation diagrams are a particular 
case of {\emph{population diagrams}} (Goldsmith and Langer [18]). 
Equation~(11) is a particular case of the more general relation:
\begin{gather}\label{eqpd:AstRus1608004Kalenskii}
\ln\frac{N_u}{g_u}=\ln\frac{3kW}{8\pi^3\nu_0S\mu^2} + \\
\nonumber{}+
\ln(C_\tau)+\ln(ff)=\ln\frac{N}{Q_{\textrm{rot}}}{-}\frac{E_u}{kT_{\textrm{rot}}},
\end{gather}
where the correction factor $C_\tau$ is a function of the column density 
and rotational temperature. The problem is reduced to searching 
the the $\chi^2$ minimum with $T_{\textrm{rot}}$, $N$, and $ff$ as free 
parameters. Population diagrams have been successfully applied by e.g., 
Gibb et al.~[19] to the analysis of the observations of various 
molecules, including methanol. Unfortunately, it is much more difficult 
to solve the nonlinear Eq.~(12) than the linear Eq.~(11).

\textbf{5.4.3. Two--temperature method.} Kalenskii et al.~[20]
developed a method for deriving source parameters from the intensities 
of the $J_0{-}J_{-1}E$ and $2_K{-}1_KE$ lines. They abandoned the LTE 
assumption and the assumption that the $J_0{-}J_{-1}E$ lines are 
optically thin, but assumed that all these lines have the same 
excitation temperature, denoted $T_{KK'}$, and that the population ratios 
for all the $J_{-1}E$ levels are described by the same temperature 
$T_{\textrm{rot}}$. Under these assumptions the expression 
for the ratio of brightness temperature of an arbitrary $J_0{-}J_{-1}E$ 
line to some reference line takes the form
\begin{gather}
\frac{T_B(J)}{T_B(4)} = \\
\nonumber{}=\frac{1-\exp [-\tau_4 S_J/S_4 \exp (-\Delta
E_{J4}/kT_{\textrm{rot}})]}{1-\exp (-\tau_4)}. \label{eq157:AstRus1608004Kalenskii}
\end{gather}
Here $T_B(J)$ and $S_J$ are the brightness temperature and the strength of 
the arbitrary line, while $T_B(4)$, $\tau_4$, and $S_4$ are the brightness 
temperature, optical depth, and strength of the $4_0{-}4_{-1}E$ line, 
chosen as the reference. When the $J_0{-}J_{-1}E$ line series
is observed, one can write the system of equations~(13) and find
$T_{\textrm{rot}}$ and $\tau_4$, solving the system by the nonlinear 
root-mean-squares method. The excitation temperature
of the $J_0{-}J_{-1}E$ transitions (denoted by Kalenskii~et~al.
as~$T_{KK'}$) can be found from the ratio of the $2_0{-}1_0E$ and 
$2_1{-}1_{-1}E$ line intensities, assuming that the lines are optically 
thin. This assumption does not contradict the initial supposition that 
the $J_0{-}J_{-1}E$ lines may not be optically thin, since SE calculations 
show that the optical depths of the $J_0{-}J_{-1}E$ lines are higher 
than those of the $2_K{-}1_K$ lines\footnote{We emphasize that 
the optical depths of the $2_K{-}1_K$ lines are of no concern when
calculating  $T_{\textrm{rot}}$ and $\tau_4$.}. Knowing $T_{KK'}$ 
and $\tau_4$, and applying the radiation transfer equation, one can find 
the brightness temperature of the reference line $T_R(4)$; then, comparing
$T_R(4)$ with the observed value, it is possible to estimate the source 
size.

Note that when the $J_0{-}J_{-1}E$ lines are optically thin,
the dependence on the optical depths vanishes from Eq.~(13). In this case 
only $T_{\textrm{rot}}$ can be determined with this method.

Thus, there are several modifications of the method of rotation diagrams,
which make it possible to determine the rotational temperature even from optically
thick lines.

\subsection{Estimation of methanol column density using a single line}
\label{singl:AstRus1608004Kalenskii}

When the gas temperature is known, molecular  column densities are often
obtained from the integrated intensity of a single, optically thin line.
For this purpose, the upper level population is first determined 
using Eq.~(10). Then, assuming that the energy levels are populated 
according to the known temperature, the molecule column density 
is calculated with
\begin{gather}
N=\frac{N_u}{g_u}\cdot Q_{\textrm{rot}}
\left(\frac{E_u}{kT_{\textrm{kin}}}\right), \label{eqcolden:AstRus1608004Kalenskii}
\end{gather}
which follows from the Boltzmann equation and is an alternative form of 
Eq.~(11).

Because the side ladders may be strongly underpopulated relative to the backbone
ladder, Eq.~(14) is not valid in the general case. Equation~(10) 
sometimes may be also wrong, especially for the $J_K{-}J_{K-1}$ lines
(see subsection~3.3). Therefore this method may yield highly
erroneous methanol column densities.

Nevertheless, the method is still used.  Hence, it is important to
understand how (un)reliable the column densities are, when determined by
this method. Here, we briefly describe some of the
pitfalls that are inherent to this method.

If there is no indication that the source density is $10^6$~cm$^{-3}$
or higher, we strongly discourage the use of results
obtained from a $J_K{-}J_{K-1}$ line, such as $J_0{-}J_{-1}E$ or 
$J_1{-}J_0E$. These lines may be weak or invisible against 
the microwave background even when the methanol column density
is high (see subsection~3.3), and Eq.~(10) may underestimate
the column densities of the upper levels by an order of magnitude
or more.

Nor does use of the $J_K{-}(J-1)_K$ lines ensure
that the column density will be determined accurately. If the energy
levels belong to the backbone ladder, then in the low-density
case  Eq.~(14) {\emph{strongly overestimates}} the methanol column density
as a result of this ladder being overpopulated
with respect to the side ladders.  If the levels belong to any
side ladder that is not adjacent to the backbone ladder (for
example, to the $K=1E$ ladder),  then in the case of low density, the methanol
column density may be {\emph{underestimated by an order of magnitude
or more}}. SE calculations show that the optimal choice is
the $J_0{-}(J-1)_0E$ line series, in addition to the $J_1{-}(J-1)_1A^{+}$ 
and~$A^{-}$ lines.   Using these line series the methanol column
density can be accurately determined to within a factor of a few.

Despite the feasibility of determining the methanol column density from 
a single line, the authors strongly recommend using the abundance of 
methanol  lines at radio frequencies to determine the
methanol column density based on observations of
several lines.

\section{SOME NOTES ON EXTERNAL RADIATION}
\label{exrad:AstRus1608004Kalenskii}

As we mentioned in the Introduction, including the effects of external 
radiation significantly complicates the analysis of methanol excitation
and is beyond the scope of this paper. The conclusions and 
recommendations presented here are valid only for sources in which
the external radiation, apart from the microwave background,
is negligible. The role of external radiation will be studied in
a dedicated paper. Nevertheless, some general notes concerning the role of 
external radiation can be made here.

It is fairly easy to analyze the role of radiation from warm (20--50~K)
dust, which is observed in the submillimeter continuum. The main effect 
of this radiation on the excitation of methanol is an increase of 
the excitation temperature of the $2\rightarrow 3$ transitions (see 
section~3). Thus, to a first approximation, this radiation affects 
methanol excitation in a way that mimics a slight increase in density. 
To account for 
the radiation of hotter sources ($>100$~K) is much more difficult and 
requires consideration of transitions between the ground and 
the excited torsional states. These transitions fall in the IR spectral 
range. However, the main effect of this radiation is again 
the redistribution of the energy level population in favor of the side 
ladders. If the radiation is sufficiently strong, it can pump Class II 
methanol masers. The strongest of these masers emit in 
the $5_1{-}6_0A^{+}$ line at 6.7~GHz and in the $2_0{-}3_{-1}E$ line 
at 12.2~GHz. Class II masers have also been observed in the $J_0{-}J_{-1}E$ lines 
at 157~GHz, repeatedly mentioned in this paper. All these lines, as well as many 
other Class II maser lines, have their upper levels on the side 
ladders and their lower levels on the backbone ladders. 

As the external radiation redistributes the energy level populations 
in favor of side ladders, one can expect that in the presence of 
radiation the distribution of energy level populations will be closer 
to LTE than in its absence. Therefore, the application of LTE methods 
to the analysis of methanol observations in, for example, hot cores, 
seems reasonable. However, it would be desirable to conduct a special 
study to determine to what degree these results are correct.  

\section{SUMMARY AND CONCLUSIONS}

We considered methanol excitation in the absence of external
radiation and analyzed the LTE methods for determining the parameters
of interstellar gas in order to understand to what extent these
methods are applicable to the exploration of real molecular clouds.

When the density is below  $10^8$~cm$^{-3}$, the rotational energy 
levels of methanol located on the side ladders are underpopulated 
relative to those on the backbone ladders; the further the ladder is from
the backbone ladder the more underpopulated it is. As a result, 
if a rotational diagram is built using the lines $J_K{-}(J-1)_K$, 
%whose upper levels havethe same quantum numbers $J$ but different quantum numbers $K$, 
e.g., the $2_K-1_K$ lines at 96~GHz, the $3_K-2_K$ lines at 145~GHz, or 
the $5_K-4_K$ lines at 241~GHz (type I rotation diagram, or RDI), 
the rotational temperature (RTI) will be much lower than the gas kinetic 
temperature. SE modeling showed that within the temperature range 
20--100~K and for methanol specific column densities no higher than 
${\sim} 10^{15}$~cm$^{-2}$/(km s$^{-1}$) the dependence of type I 
rotational temperatures on these parameters is weak. Within these
parameters  
RTIs depend on density and can be used to  estimate this parameter.
At a density about $10^4{-}10^5$~cm$^{-3}$ RTIs fall in the range 2--5~K, 
increasing slightly with methanol column density. At a density about 
$10^6$~cm$^{-3}$ RTI are from 6--8~K; RTIs of about  11~K or 
higher show that the density is at least ${\sim} 10^7$~cm$^{-3}$.

SE calculations show that  over a wide range of densities, typical 
for  molecular clouds in the Galaxy, the ratios of level populations 
within the same ladder are described by a single temperature, which 
is much closer to the kinetic temperature than the type I rotational 
temperature. From this it follows that the temperature should be 
determined using rotation diagrams built from lines whose upper levels 
are located in the same ladder (type II rotation diagrams, or RDII). 
Suitable lines are e.g., the $J_0{-}J_{-1}E$ lines at 157~GHz, 
the $J_1{-}J_0E$ lines at 165~GHz or the $J_2{-}J_1E$ lines at 25~GHz.
However, even RTIIs accurately reproduce the kinetic temperature only when 
the density is about $10^7$~cm$^{-3}$ or higher. When the density 
is about $10^6$~cm$^{-3}$ or lower, the type II rotational temperatures, 
being higher than the type I  temperatures, are nevertheless notably
lower than the kinetic temperature.

It is desirable to determine the kinetic temperature with rotational
diagrams using the following approach. One should build two rotation 
diagrams: RDI and RDII. First, one should estimate the gas 
density using RDI. If the density is about $10^7$~cm$^{-3}$
or higher, RTII accurately reflect the kinetic temperature. If the density
is about  $10^6$~cm$^{-3}$, the kinetic temperature can be estimated,
multiplying RTII by a factor of 1.5--2. If the density is about 
$10^5$~cm$^{-3}$ or lower, RTII will be lower than the kinetic temperature 
by a factor of three or more and can be used only as a lower limit 
on the kinetic temperature.

Methanol column densities, determined with a rotation diagram, also can 
be erroneous and should be used with caution. When the column density is determined 
with an RD96 or RD145, the true value may be underestimated by a factor 
of 2--5, depending on the density. When this parameter is determined 
using any RDII or RD241, the accuracy depends on the source density. 
When the density is lower than $10^6$~cm$^{-3}$, the accuracy 
is very low: the ratio of the true to the derived value may vary
within the range from 0.01--0.003 in the case of RDII to ${\sim} 150$
in the case of RD241. When the density is about $10^6$~cm$^{-3}$, RDIIs 
permit the determination of methanol column density with an accuracy 
of about a factor of five, and DR241 overestimates it by a 
factor of 15--20.   When the density 
is about $10^7$~cm$^{-3}$ or higher RD241 determines column density with 
an an accuracy to a factor of about 1.5 or better, and RDIIs provide 
an accuracy within  about a factor of two.

When the gas temperature is known, the molecular column density can be estimated
from the intensity of a single line, assuming LTE.  For methanol that is not 
in LTE, the error of a column density determined by a single line can be
greater than an order of magnitude.   Thus, the authors strongly 
recommend using the large number of methanol  lines available
at radio frequencies to determine the methanol column density based on 
the observations of several lines. When the $J_0{-}(J-1)_0E$ or 
$J_1{-}(J-1)_1A^{+}$ and~$A^{-}$ lines are used to determine the
column density, the relative error of the derived value will be no larger 
than several units.

The authors are grateful to the referee for useful comments. The work
was partially supported by the UNAM DGAPA project PAPIIT-IN114514.

\newpage

\begin{figure*}[t!]
%%% Figure:1
\begin{center}
\includegraphics[width=0.7\textwidth]{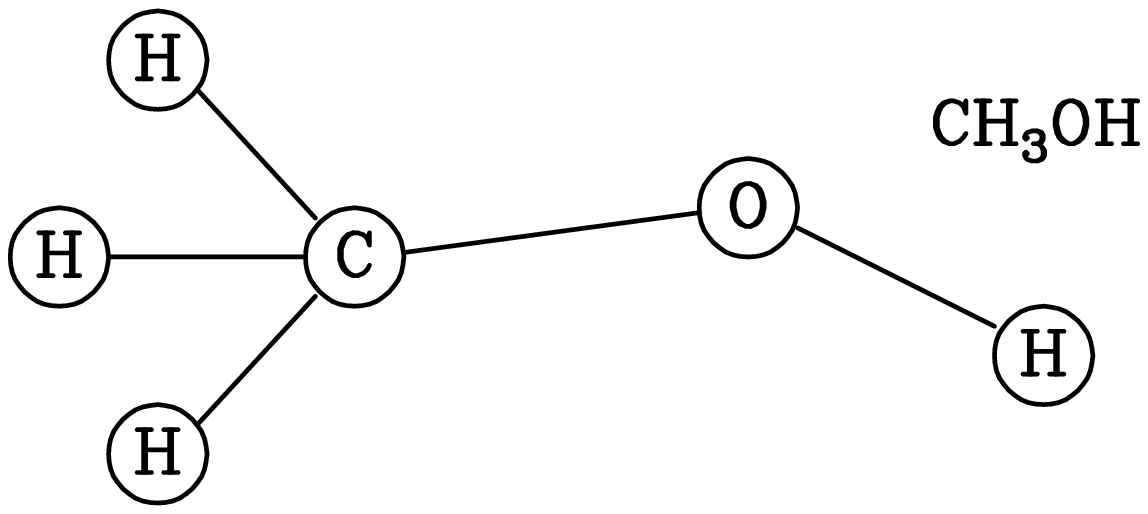}
\end{center}
\caption{The structure of methanol molecule. \hfill}
\end{figure*}

\begin{figure*}[t!]
%%% Figure:2
\includegraphics[angle=-90, width=0.9\textwidth]{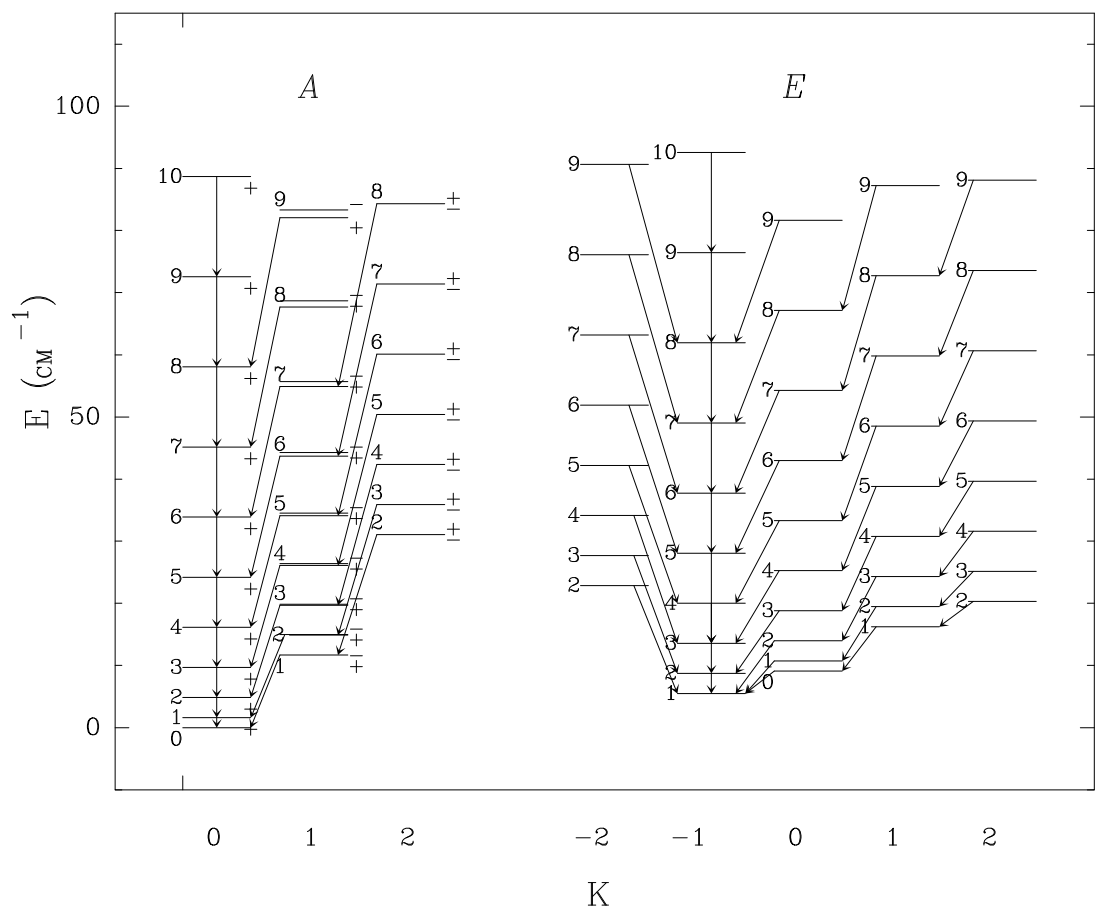}
\caption{Energy levels of $A$ and $E$ methanol. Arrows denote the fastest
spontaneous transitions from each level. The arrows that show such 
transitions from the $J_2A^{+}$ and $J_2A^{-}$ levels are indistinguishable 
at the scale of this picture. \hfill}
\end{figure*}

\begin{figure}[t!]
%%% Figure:3
\begin{center}
\includegraphics{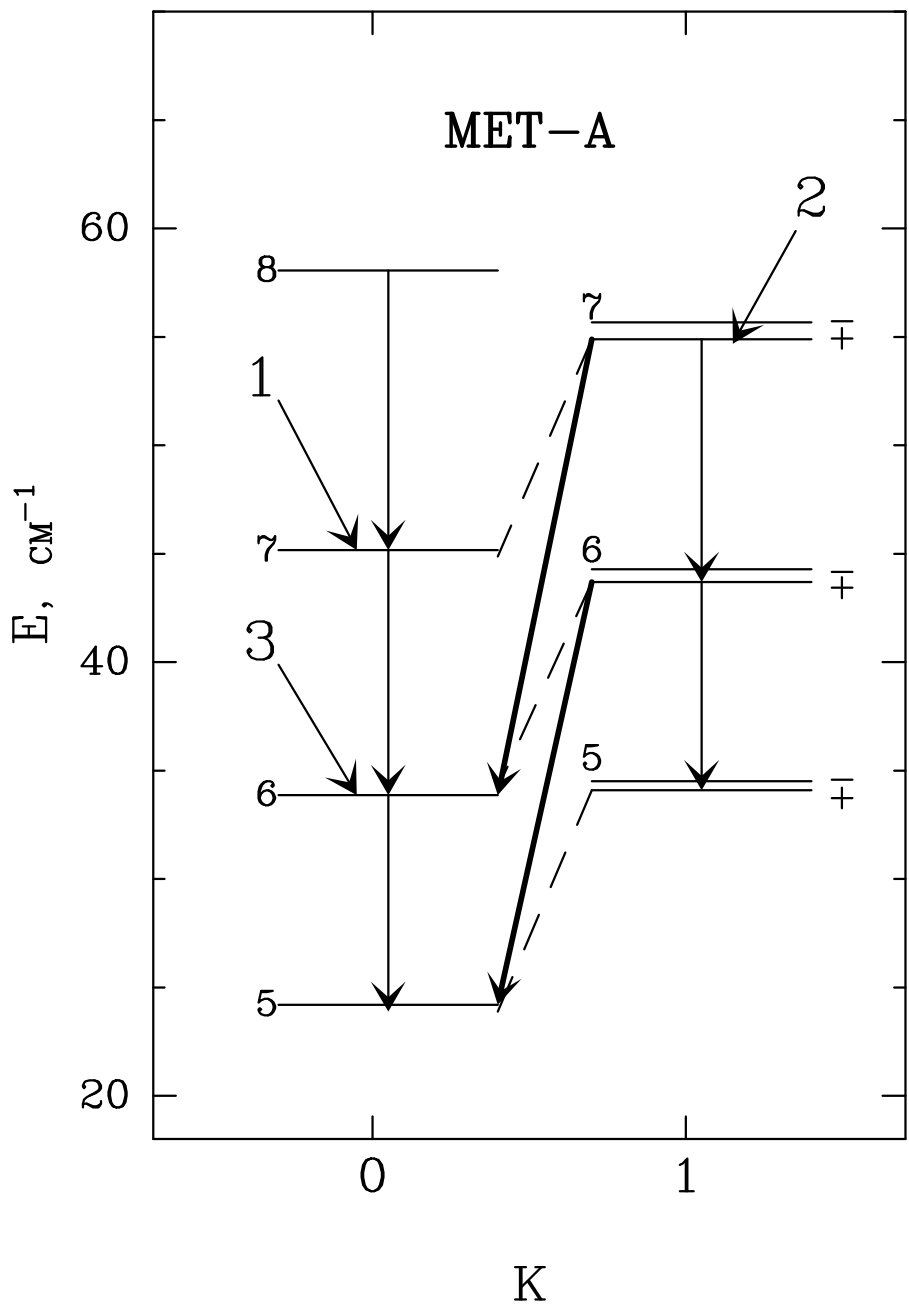}
\end{center}
\caption{Selected energy levels of $A$-methanol. Thick arrows show  transitions
that empty the $J_1A^{+}$ levels. Dashed lines show paths between 
the $J_1A^{+}$ and $J_0A^{+}$ levels; radiative transitions between these levels
are prohibited by selection rules.
 \hfill}
\end{figure}

\begin{figure*}[t!]
%%% Figure:4
\includegraphics[width=0.9\textwidth]{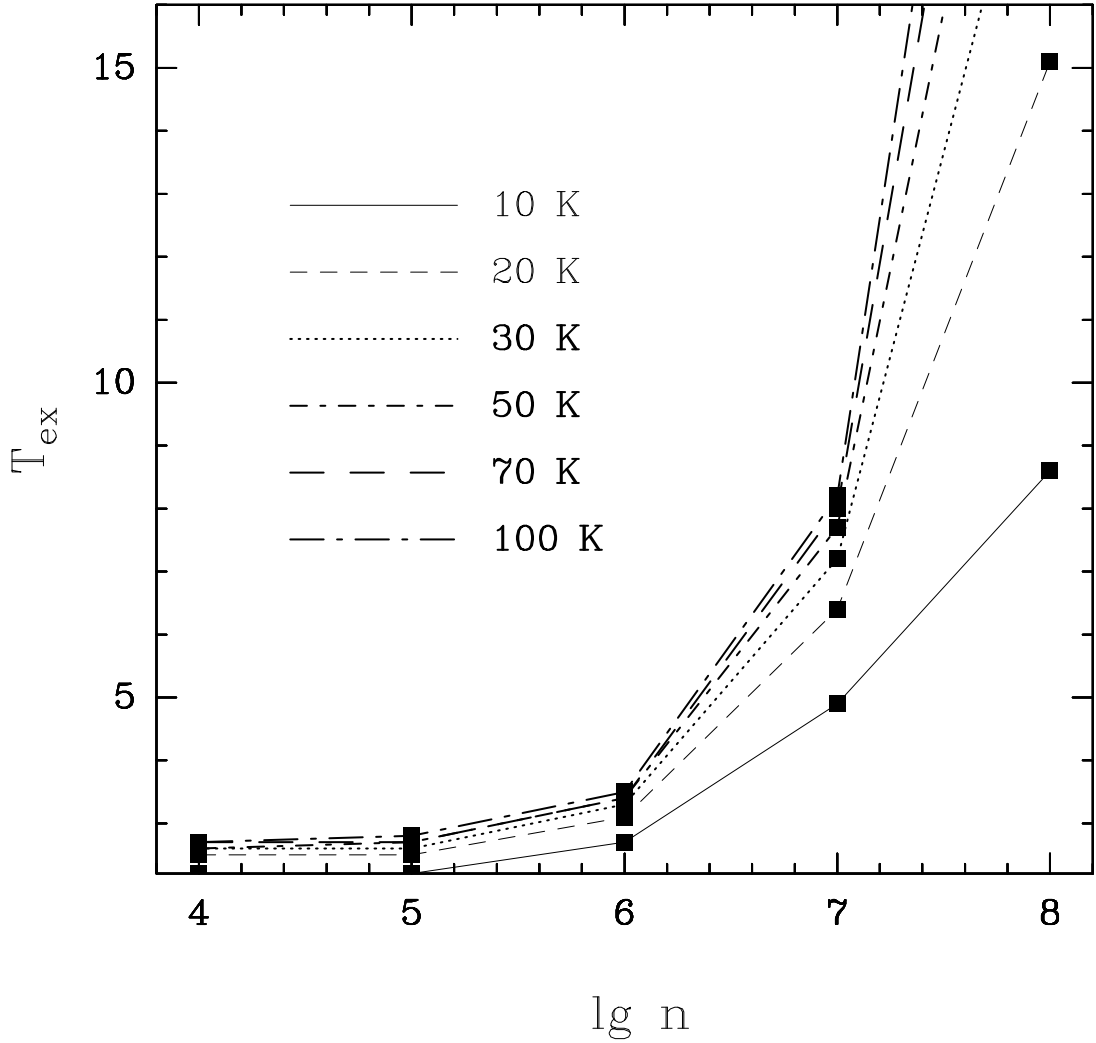}
\caption{The density dependence of the excitation temperature
of the $5_{-1}-5_0E$ transition, which is the $2\rightarrow 1$ transition
in the system of the $5_{-1}E$, $5_0E$, and $4_{-1}E$ levels. 
The dependence is calculated for six kinetic temperatures
in the range 10--100~K.
\hfill}
\end{figure*}

\begin{figure*}[t!]
%%% Figure:5
\includegraphics[width=0.9\textwidth]{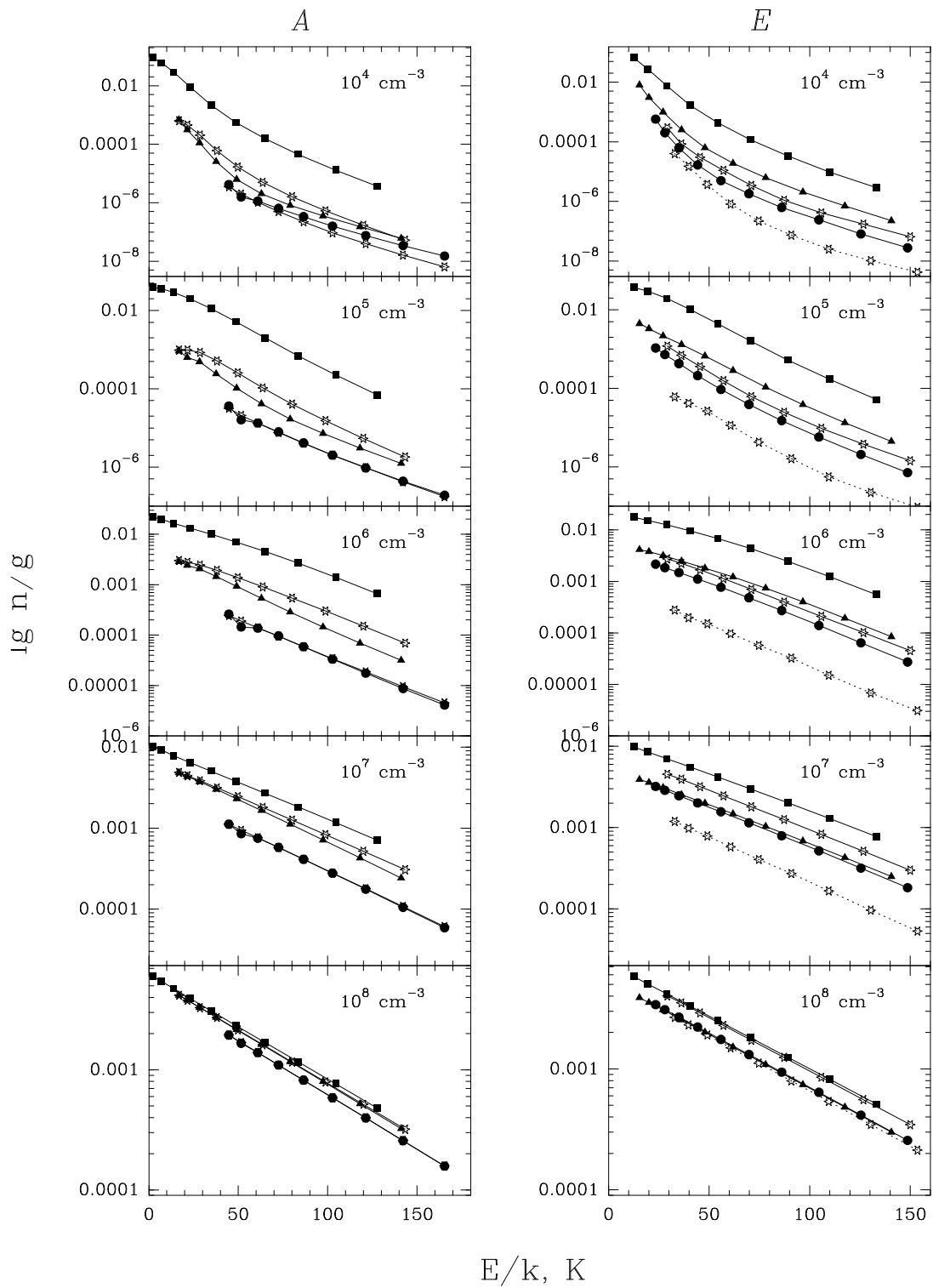}
\caption{Results of SE calculations. Left column: methanol $A$. Filled squares 
denote the $J_0A^{+}$ levels; triangles, the $J_1A^{+}$ levels; asterisks, 
the $J_1A^{-}$ levels; filled circles, the $J_2A^{+}$ levels; open 
stars, the $J_2A^{-}$ levels. Right column: methanol $E$. Filled 
squares denote the $J_{-1}E$ levels; triangles, the $J_0E$ levels;
filled circles, the $J_1E$ levels; asterisks, connected by solid lines, 
the $J_2E$ levels; asterisks, connected by dotted lines, the $J_{-2}E$ 
levels. \hfill}
\end{figure*}

\begin{figure*}[t!]
%%% Figure:6
\includegraphics[width=0.9\textwidth]{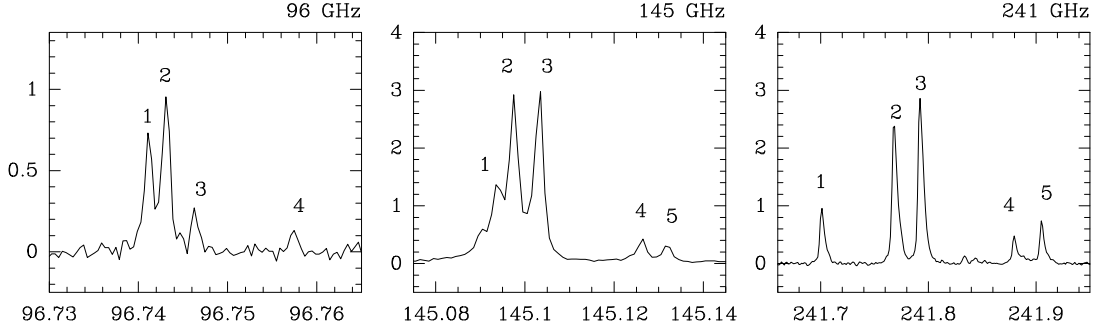}
\caption{Examples of lines at 96, 145, and 241~GHz. The notation is
as follows: 96~GHz: the $2_{-1}{-}1_{-1}E$ line ({\textit{1}}); 
the $2_0{-}1_0A^{+}$ line ({\textit{2}}); the $2_0{-}1_0E$ line 
({\textit{3}}); the $2_1{-}1_1E$ line ({\textit{4}}); 145~GHz: 
the $3_0{-}2_0E$ line ({\textit{1}}); the $3_{-1}{-}2_{-1}E$ line
({\textit{2}}); the $3_0{-}2_0A^{+}$ line ({\textit{3}}); the blend
of the $3_2{-}2_2E$ and $3_{-2}{-}2_{-2}E$ lines ({\textit{4}}); 
the $3_1{-}2_1E$ line ({\textit{5}}); 241~GHz: the $5_0{-}4_0E$
line ({\textit{1}}); the $5_{-1}{-}4_{-1}E$ line ({\textit{2}}); 
the $5_0{-}4_0A^{+}$ line ({\textit{3}}); the $5_1{-}4_1E$ line
({\textit{4}}); the blend of the $5_2{-}4_2E$ and 
$5_{-2}{-}4_{-2}E$ lines ({\textit{5}}). The bump left of the
$3_0{-}2_0E$  transition (middle plot) is a c-C$_3$H$_2$ line. \hfill}
\end{figure*}

\begin{figure*}[t!]
%%% Figure:7
\includegraphics[width=0.9\textwidth]{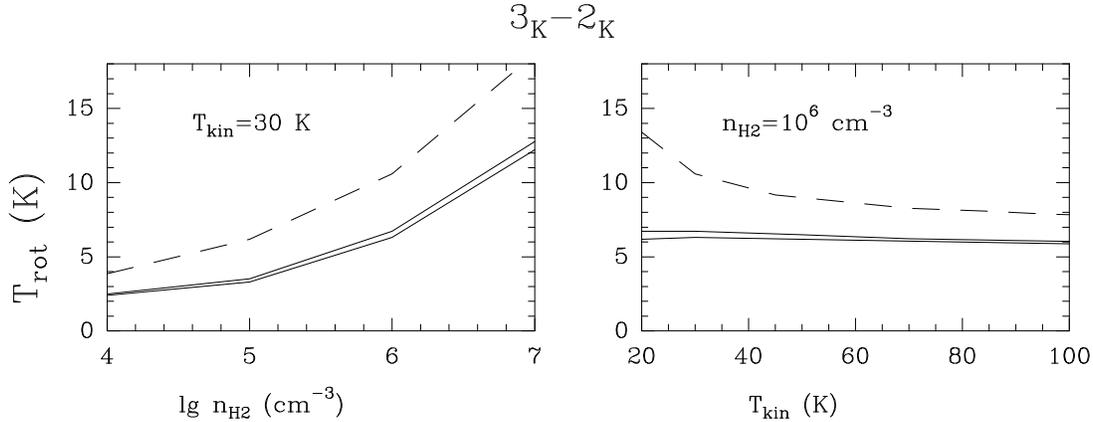}
\caption{Dependences of RT145 (i.e., based on the 145 GHz lines) on density (left) and on
kinetic temperature (right). Solid lines on both plots show
these dependences for methanol column densities of
$N_{\textrm{CH$_3$OH}}/dV=2\times 10^{13}$ and $2\times
10^{14}$~cm$^{-2}$/(km~s$^{-1}$), while the dashed line is for the column 
density $N_{\textrm{CH$_3$OH}}/dV=2\times
10^{15}$~cm$^{-2}$/(km~s$^{-1}$)). \hfill}
\end{figure*}

\begin{figure*}[t!]
%%% Figure:8
\begin{center}
\includegraphics{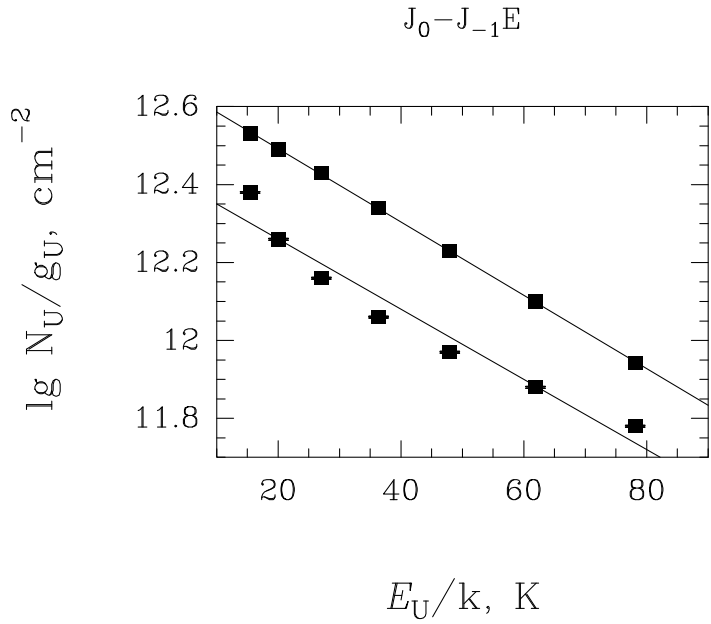}
\end{center}
 \caption{Model rotation diagrams built from the $J_0{-}J_{-1}E$ 
lines at 157~GHz for different specific column densities of methanol. 
The upper diagram:
$N_{\textrm{CH$_3$OH}}/dV=2\times
10^{13}$~cm$^{-2}$/(km~s$^{-1}$) (optically thin lines). The line has 
been shifted upwards by a factor of about two for ease of comparison
with the second diagram. The lower diagram:
$N_{\textrm{CH$_3$OH}}/dV=2\times 10^{15}$~cm$^{-2}$/(km~s$^{-1}$)
(optical depths of the lines are greater than unity). \hfill}
\end{figure*}

\begin{table*}[t!]
%%% Table:1
\caption{The properties of type I and II rotation diagrams. Models used 
for the building of rotation diagrams were computed by the LVG method.
The even columns from 2 to 12 present rotational temperatures for 
different models with $T_{\textrm{kin}}=50$~K and for different line samples. 
The odd columns (from 3 to 13) present the ratios of methanol 
column densities obtained from the rotational diagrams ($N_{\textrm{rot}}$)
to the model ($N_{\textrm{mod}}$) column densities. The dashes mean that 
the corresponding lines are not seen because their excitation temperatures 
are close to the microwave background temperature. The correct 
parameter values are presented to the right of the slashes (/); the values to the
left of slashes are derived using line blends (see section~5.2) and 
do not reproduce the populations of the upper levels of the relevant lines.}
\scriptsize
\begin{tabular}{|c|c|c|c|c|c|c|c|c|c|c|c|c|}
\noalign{\medskip}
\hline
\noalign{\smallskip}
%\tiny
           &\multicolumn{6}{c|}{Type I}                     &\multicolumn{6}{c|}{Type II}\\
\hline
           &\multicolumn{2}{c|}{96~GHz}
&\multicolumn{2}{c|}{145~GHz}
&\multicolumn{2}{c|}{241~GHz}
                          &\multicolumn{2}{c|}{157~GHz}
                            &\multicolumn{2}{c|}{165~GHz}
                                              &\multicolumn{2}{c|}{25~GHz} \\
\hline
$n_{H_2}$  &$T_{rot}$&$\frac{N_{rot}}{N_{mod}}$
                         &$T_{rot}$&$\frac{N_{rot}}{N_{mod}}$ 
                                       &$T_{rot}$&$\frac{N_{rot}}{N_{mod}}$
                                                       &$T_{rot}$&$\frac{N_{rot}}{N_{mod}}$
                                                                       &$T_{rot}$&$\frac{N_{rot}}{N_{mod}}$  
                                                                                        &$T_{rot}$&$\frac{N_{rot}}{N_{mod}}$ \\
(cm$^{-3}$)& (K)  &      & (K)   &     & (K)    &      & (K)     &     & (K)     &      & (K)     &      \\
\hline
\multicolumn{13}{|c|}{$N_{mod}/dV=2\times 10^{13}$~см$^{-2}$/(км с$^{-1}$)}\\
\hline
 $10^4$   & 2.3  & 0.7 & 2.7   & 1.7   &  3.4   &39$^4$    & --     & --   &  11$^1$ & 0.003& 12$^{1,3}$&0.003\\
 $10^5$   & 3.5  & 0.4 & 3.7   & 1.2   &  4.2   &5.8$^4$   & --     & --   &  14$^2$ & 0.05 & 16$^{2,3}$&0.17 \\
 $10^6$   & 6.0  & 0.2 &7.1/5.9&0.3/0.4& 7.9/5.6&0.4$^4$/18& 32     & 0.4  &  30     & 0.3  & 30$^3$    & 0.7 \\
 $10^7$   & 11.4 & 0.2 &18/11  &0.3/0.2& 23/11  &1.0/1.4   & 46     & 0.8  &  45     & 0.7  & 45$^3$    & 2   \\
 $10^8$   & 23.0 & 0.4 &105/24 &2.3/0.3& 220/23 &7.9/0.7   & 49     & 0.9  &  49     & 0.8  & 50$^3$    & 1.7 \\
\hline
\multicolumn{13}{|c|}{$N_{mod}/dV=2\times 10^{14}$~cm$^{-2}$/(km s$^{-1}$)}\\
\hline
 $10^4$   & 2.5 & 0.5 &2.8/2.5&1.5$^4$/2.1$^4$& 3.4/2.6 &92$^4$/92$^4$ & -- & --   & 11      & 0.003& 12      & 0.03 \\
 $10^5$   & 3.6 & 0.4 &4.0/3.4&0.9$^4$/1.3$^4$& 4.3/3.2 &44$^4$/56$^4$ & 11 & 0.04 & 14      & 0.05 & 16      & 0.17 \\
 $10^6$   & 6.2 & 0.2 &7.5/6.1&0.4/0.4        & 8.3/5.7 &3.3$^4$/13$^4$& 34 & 0.4  & 30      & 0.3  & 29      & 0.7  \\
 $10^7$   & 12  & 0.2 & 20/12 &0.4/0.2        & 24/12   &0.9$^4$/1.3   & 46 & 0.8  & 45      & 0.7  & 44      & 2.0 \\
 $10^8$   & 23  & 0.3 &104/23 &2.0/0.3        &401/24   & 20$^4$/0.7   & 49 & 0.9  & 49      & 0.9  & 50      & 1.7 \\
\hline
\multicolumn{13}{|c|}{$N_{mod}/dV=2\times 10^{15}$~cm$^{-2}$/(km s$^{-1}$)}\\
\hline
 $10^4$   & 3.6 & 0.2 &4.0/3.6&0.2$^4$/0.3$^4$& 3.9/3.1 &16$^4$/142$^4$&7.5 & 0.04 & -- & --   & 14      & 0.01  \\
 $10^5$   & 5.1 & 0.3 &5.8/5.4& 0.4/0.4       & 6.4/5.3 &4.1$^4$/10$^4$& 28 & 0.15 & 14 & 0.04 & 16      & 0.15  \\
 $10^6$   & 8.1 & 0.2 &11/8.7 & 0.3/0.2       & 13/9.0  &1.0/2.0       & 47 & 0.4  & 32 & 0.2  & 29      & 0.7  \\
 $10^7$   & 14  & 0.2 &30/15  & 0.4/0.2       & 48/16   &1.0/0.7       & 48 & 0.5  & 45 & 0.6  & 45      & 2.1  \\
 $10^8$   & 26  & 0.4 &242/28 & 8/0.3         &-440/30  &$\infty$/0.6  & 49 & 0.7  & 49 & 0.8  & 51      & 1.5  \\ 
\hline
\end{tabular}

\bigskip
$^1$the line brightness temperatures are about or below 0.001~K; \\
$^2$the line brightness temperatures are about or below 0.01~K; \\
$^3$the excitation temperatures of some lines are negative; \\
$^4$the error is larger than the value itself.\hfill
%\caption{\footnotesize{$^{*}$~­®áª . \hfill}}
%\addtocounter{table}{-1}
\end{table*}

\end{document}